\documentclass{article}

\usepackage{iclr_aims}
\usepackage{times}
\usepackage{microtype}
\usepackage{hyperref}
\usepackage{amsmath}
\usepackage{amssymb}
\usepackage{xcolor}
\usepackage{graphicx}
\usepackage{booktabs}
\usepackage{subcaption}
\usepackage{comment}

\iclrfinalcopy

\newcommand{\BorderlinePct}{30\%}
\newcommand{\BorderlineFrac}{0.3}

\newcommand{\EstRho}{0.41}
\newcommand{\EstDelta}{0.024}
\newcommand{\EstDeltaCiLow}{0.003}
\newcommand{\EstDeltaCiHigh}{0.045}
\newcommand{\EstNetImpact}{24}
\newcommand{\EstImpactN}{30{,}000}
\newcommand{\EstImpactS}{0.3}
\newcommand{\EstRhoPapers}{1000}
\newcommand{\EstRhoRounds}{40}
\newcommand{\EstRhoModel}{GPT-5-mini}
\newcommand{\EstRhoCostDollars}{120}
\newcommand{\AblationPapers}{1000}
\newcommand{\AblationRounds}{40}

\newcommand{\AblationMarginalFig}{figures/fig_marginal_fraction_1020.pdf}
\newcommand{\AblationMarginalRhoFig}{figures/fig_marginal_fraction_rho_1020.pdf}
\newcommand{\AblationCenteringFig}{figures/fig_acceptance_centering_1020.pdf}
\newcommand{\AblationDeltaWidthFig}{figures/fig_delta_width_1020.pdf}

\newcommand{\AucTablePath}{tables/auc_table_1020.tex}

\newcommand{\CostBatchDiscount}{50\%}

\newcommand{\CostScaledDollars}{3{,}600}
\newcommand{\CostScaledDollarsAfterDiscount}{1{,}800}

\newcommand{\TargetAcceptanceRateFrac}{0.25}

\begin{document}

\title{Allocate Marginal Reviews to Borderline Papers Using LLM Comparative Ranking}
\author{
  Elliot L. Epstein$^{1}$, Rajat Dwaraknath$^{1}$, John Winnicki$^{1}$, Thanawat Sornwanee$^{1}$ \\
  $^{1}$Stanford University, Stanford, CA 94305, USA \\
  \texttt{\{epsteine, rajatvd, winnicki, tsornwanee\}@stanford.edu}
}
\date{}
\maketitle
\makeatletter
\fancyhead{}
\renewcommand{\headrulewidth}{0pt}
\makeatother

\begin{abstract}
This paper argues that large ML conferences should allocate marginal review capacity primarily to papers near the acceptance boundary, rather than spreading extra reviews via random or affinity-driven heuristics.
We propose using LLM-based comparative ranking (via pairwise comparisons and a Bradley--Terry model) to identify a borderline band \emph{before} human reviewing and to allocate \emph{marginal} reviewer capacity at assignment time.
Concretely, given a venue-specific minimum review target (e.g., 3 or 4), we use this signal to decide which papers receive one additional review (e.g., a 4th or 5th), without conditioning on any human reviews and without using LLM outputs for accept/reject.
We provide a simple expected-impact calculation in terms of (i) the overlap between the predicted and true borderline sets ($\rho$) and (ii) the incremental value of an extra review near the boundary ($\Delta$), and we provide retrospective proxies to estimate these quantities.
\end{abstract}

\section{Introduction}
Conferences typically aim to meet a minimum number of reviews per paper.
In practice, there is often some surplus reviewer capacity beyond that minimum.
Public review corpora report average reviews per paper above three in several venues and years, implying a marginal surplus of review slots in practice \citep{10.1145/3746252.3761506,Plank2019CiteTrackedAL,Su17072025}.
The natural question is where those marginal reviews should go to improve decisions the most.

The marginal value of an extra review is typically highest near the acceptance boundary, where score variance is high and decisions are most sensitive to reviewer noise.
It is lowest for papers that are clearly strong or clearly weak.
In practice, surplus capacity is typically absorbed by load balancing and affinity objectives \citep{kobren2019papermatchinglocalfairness,charlin2012frameworkoptimizingpapermatching,Charlin2013TheTP}.
Randomized assignment has also been proposed to mitigate manipulation in reviewer matching \citep{NEURIPS2020_93fb3947}.

\begin{figure*}[t]
  \centering
  \includegraphics[width=0.95\textwidth]{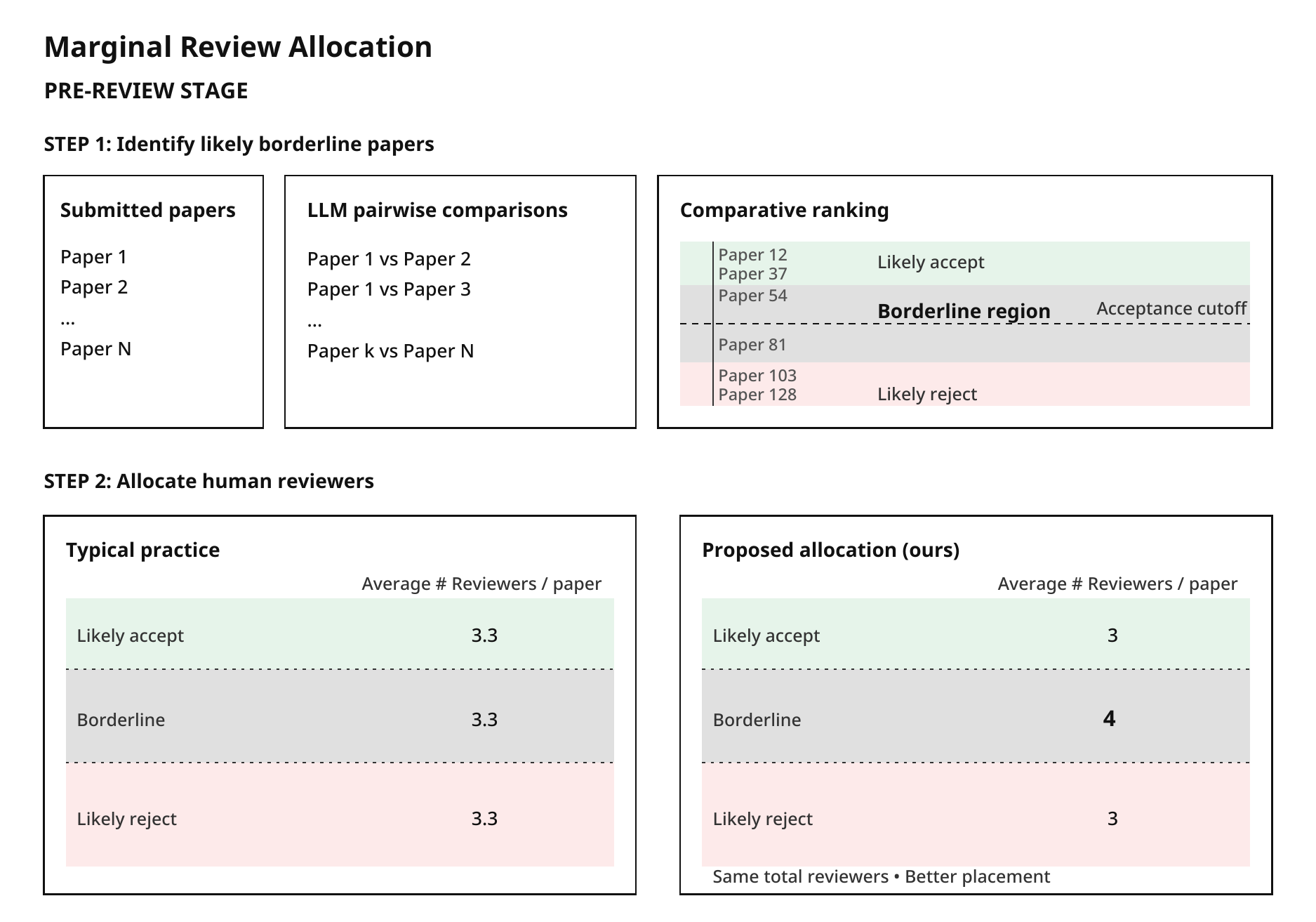}
  \caption{Schematic of the marginal review allocation pipeline.
Step 1 uses LLM pairwise comparisons to construct a comparative ranking and identify a borderline band around the acceptance percentile.
  Step 2 allocates marginal reviews to that band while keeping total reviewer load fixed, with numbers shown for illustration.}
  \label{fig:marginal_review_allocation}
\end{figure*}

Why has this been difficult to implement?
Identifying borderline papers typically requires early human reading, which arrives too late to inform reviewer-count decisions at assignment time.
By the time human signal accumulates, the review process is already underway and marginal review allocation is harder to adjust.

Recent long-context LLMs~\citep{geminiteam2025geminifamilyhighlycapable} make a lightweight pre-review triage pass feasible, enabling reviewer-count decisions before human reviewing begins.
We do not need calibrated absolute scores.
We only need a rough ranking that separates likely borderline papers from the rest.
Pairwise comparisons are a good fit because comparative judgments are less sensitive to calibration drift than absolute scores and can be aggregated into a robust ordering.
This motivates using LLMs not to decide outcomes, but to prioritize where human effort is most valuable.

Given this, we argue that
\textbf{conferences should allocate marginal review capacity to papers near the acceptance boundary, using LLM-based comparative ranking only to target human effort, while keeping accept/reject decisions fully human and keeping the LLM signal hidden from decision-makers.}
Figure~\ref{fig:marginal_review_allocation} summarizes the pipeline.
This intervention happens before review begins, so it changes only reviewer-count allocation at assignment time (not review content, rebuttals, or decision procedures).
It fits within existing workflows where area chairs can adjust assignments after the initial match.
It is a minimal change to existing workflows at NeurIPS, ICML, and ICLR, does not introduce a two-stage review that could delay timelines, and remains separate from the final decision process.

Our contributions are threefold.
First, we advance a position: LLMs should be used to allocate scarce human reviewer effort to identify where marginal reviews matter most, rather than to automate accept/reject decisions.
Second, we formalize marginal review allocation as a policy choice by defining a borderline band from surplus capacity and deriving expected net improved paper decisions in terms of $\rho$ (borderline overlap fraction) and $\Delta$ (marginal benefit of an extra review), and we outline a minimal-change pipeline that uses LLM pairwise comparisons with Bradley--Terry ranking to construct the band early while keeping final decisions fully human.
Third, we provide initial evidence from 1,000 papers at ICLR 2025 and probe robustness to band fraction and centering.

The paper is organized as follows.
Section~\ref{sec:llm-ranking} describes the LLM comparative ranking pipeline and how we fit the Bradley--Terry model.
Section~\ref{sec:lit-review} situates our proposal in the literature on reviewer assignment and LLM-based reviewing.
Section~\ref{sec:impact} formalizes expected net improved paper decisions, and Section~\ref{sec:estimation} details how we estimate $\rho$ and $\Delta$.
Section~\ref{sec:results} presents empirical validation and ablations, followed by a discussion in Section~\ref{sec:discussion} and alternative views in Section~\ref{sec:alternative-views}.
We close with a conclusion in Section~\ref{sec:conclusion}.

\section{LLM Comparative Ranking}
\label{sec:llm-ranking}
We describe one concrete pipeline to make the proposed allocation policy operational; the position itself does not depend on this specific modeling choice.
Conceptually, we only require a coarse ordering that separates likely borderline papers from clearly strong or weak ones.
We estimate paper quality with pairwise LLM comparisons and a Bradley-Terry model \citep{bradley1952rank,zhang2025from}.
For paper $i$ with latent score $\theta_i$, the win probability against paper $j$ is
\[
P(i \succ j) = \frac{\exp(\theta_i)}{\exp(\theta_i) + \exp(\theta_j)}.
\]
Given observed outcomes from many pairwise matches, we fit $\{\theta_i\}$ by maximum likelihood and rank papers by $\theta_i$.
Equivalently, for each observed match $(i,j)$ with outcome $y_{ij}\in\{0,1\}$ indicating whether $i$ wins, the log-likelihood is
\[
\sum_{(i,j)} \left[ y_{ij}\log P(i \succ j) + (1-y_{ij})\log\bigl(1-P(i \succ j)\bigr) \right],
\]
which we maximize over $\{\theta_i\}$ subject to an identifiability constraint (e.g., $\sum_i \theta_i=0$).
The resulting scores are unique up to a constant shift, and their ordering is the ranking.

\paragraph{Match system.}
We run multiple rounds of random pairings across the submission set, recording a binary winner for each match.
These outcomes define the Bradley--Terry likelihood; the fitted scores yield a total ordering used to define the borderline band.

\paragraph{Prompt and parsing.}
Each comparison uses a structured prompt with two papers and asks the LLM to choose one.
The prompt includes title, abstract, figure/table captions (when available), and main text, and requests a JSON output with a single chosen paper.
We parse the JSON field (``paper\_1'' or ``paper\_2'') and ignore all other content.

\paragraph{Why pairwise comparisons.}
Pairwise comparisons are attractive in this setting because comparative judgments are less sensitive to scale drift and calibration than pointwise scores, and they can be aggregated into a total order with standard models such as Bradley-Terry.
This makes the approach robust to model upgrades and prompt changes, provided the pairwise preferences are reasonably consistent.


\paragraph{PDF extraction.}
We extract text from PDFs with ScienceBeam, following \citet{zhang2025from}.
We truncate inputs to the first $P_{\max}=10$ pages before parsing to reduce length bias and cost.

\section{Related Work}
\label{sec:lit-review}

\subsection*{Reviewer assignment and conference operations}
Large conferences rely on automated reviewer assignment based on reviewer bids, text similarity, subject areas, and conflict constraints, typically solved as an optimization problem under load and coverage constraints \citep{charlin2012frameworkoptimizingpapermatching,Charlin2013TheTP,10.1145/2645710.2645749,leytonbrown2022matchingpapersreviewerslarge,LEYTONBROWN2024104119}.
Fairness- and accuracy-aware objectives have been formalized for assignment, including max--min fairness and statistical accuracy guarantees \citep{stelmakh2019peerreview4allfairaccuratereviewer}.
Randomized assignment has been proposed to mitigate manipulation, preserve anonymity, and enable counterfactual evaluation of matching policies \citep{NEURIPS2020_93fb3947,xu2023a,saveski2023counterfactual}.
Large Conference Matching (LCM) integrates data aggregation, constrained optimization, and a two-phase reviewing process (often described as a 2+2 scheme) that shifts reviewer resources toward papers near the decision boundary after initial human reviews are in \citep{leytonbrown2022matchingpapersreviewerslarge,LEYTONBROWN2024104119}.
Our proposal is complementary but differs in timing: we allocate marginal reviews \emph{before} any human has read the papers, using LLM comparative ranking to pre-identify the likely borderline band once a minimum coverage target is met.
More broadly, prior allocation methods are typically framed around meeting a minimum reviewer count and optimizing assignment quality, but they are less explicit about how to deploy \emph{marginal} reviewers when total capacity exceeds the minimum.
For example, classic assignment systems such as the Toronto Paper Matching System formalize coverage and load constraints but do not, to our knowledge, specify a dedicated policy for surplus capacity beyond those constraints \citep{Charlin2013TheTP}.
In practice, surplus capacity is typically absorbed by load-balancing and affinity-based assignment objectives \citep{Charlin2013TheTP,10.1145/2645710.2645749}, whereas our focus is to explicitly target that surplus to the likely borderline region.

Public review corpora provide evidence that mean review counts often exceed three, implying a marginal surplus of review slots beyond a three-review baseline.
RottenReviews reports average reviews per paper of 4.47 (NeurIPS 2024) and 3.86 (ICLR 2024) \citep{10.1145/3746252.3761506}, and CiteTracked reports NeurIPS averages above three across 2013--2018, with some years substantially higher \citep{Plank2019CiteTrackedAL}.
The ICML 2023 ranking experiment reports 3.08 reviews per submission pre-rebuttal and 3.29 post-rebuttal for its ranked-submission subset \citep{Su17072025}.
These figures vary by venue, year, and subset, but they indicate that marginal review capacity exists in practice, which creates an allocation question about where to place those extra reviews.

\subsection*{Decision variability and the marginal value of extra reviews}
Peer-review outcomes exhibit substantial variability and decision noise near the acceptance boundary, as shown by analyses of NeurIPS 2016 (NIPS 2016) and related Bayesian estimates of arbitrariness \citep{shah2018designanalysisnips2016,francois2015arbitrarinesspeerreviewbayesian}.
Mechanism-design and calibration perspectives further highlight how noisy ratings and strategic behavior can persist under structured review policies \citep{lu2023calibrating,srinivasan2023auctionspeerpredictionacademic}.
Recent work leverages author-provided rankings to calibrate scores and improve decision reliability, including the Isotonic Mechanism and the ICML 2023 ranking experiment \citep{su2021you,Su17072025}.
These findings motivate explicitly modeling the marginal benefit of additional reviewers rather than treating review capacity as fixed.

\subsection*{LLMs for reviewing and automated quality estimation}
Most LLM-focused work treats LLMs as substitutes or assistants for human reviewers, evaluating their ability to generate reviews, predict scores with calibrated uncertainty estimates~\cite{epstein2026llms}, or critique papers \citep{liu2023reviewergptexploratorystudyusing,robertson2023gpt4slightlyhelpfulpeerreview,idahl2025openreviewerspecializedlargelanguage,zhu-etal-2025-deepreview,zhou-etal-2024-llm,liang2023largelanguagemodelsprovide,zhang2025reviewingscientificpaperscritical}.
Recent conference pilots and reports, including a AAAI program, emphasize AI assistance that provides factual review content without scores, leaving accept or reject decisions entirely to humans \citep{aaai2026llmpressrelease}.
Our proposal is complementary and could be deployed alongside such systems, since we use LLMs only to target marginal reviewer allocation rather than to generate reviews.
Risk analyses emphasize susceptibility to manipulation, bias, and unreliable judgments under long-context or incomplete inputs \citep{ye2024yetrevealingrisksutilizing,akella2025prereviewpeerreviewpitfalls}.
Surveys synthesize the growing literature on automated scholarly paper review and its limitations \citep{10.1016/j.inffus.2025.103332}.
Parallel lines of work explore pairwise or debiased quality estimation, including LLM-based pairwise comparisons and scalable pairwise training with pointwise inference \citep{zhang2025from,zhao2025naipv2debiasedpairwiselearning}.
NAIPv2 in particular trains on pairwise preferences but performs pointwise inference at deployment, which could reduce latency and cost compared to full pairwise aggregation, and we do not yet compare against that alternative.

\subsection*{Gap: allocating human review resources with LLM signal}
Despite extensive work on assignment and on LLMs as reviewers, there is limited focus on using LLM-derived comparative signals to steer the allocation of human review effort.
Existing pipelines such as LCM shift resources toward borderline papers but do not leverage LLM comparative ranking to identify those papers \citep{leytonbrown2022matchingpapersreviewerslarge,LEYTONBROWN2024104119}.
Our work targets this gap by using LLM pairwise ranking to estimate the borderline band and by quantifying the marginal value of reallocating human reviews to that band.

\section{Expected Net Improved Paper Decisions}
\label{sec:impact}
Assume $N$ submissions and a mean surplus of $s$ reviews per paper beyond the minimum $r_{\min}$, so the venue has $sN$ ``+1 review'' slots available.
If each paper can receive at most one additional review, then at most $sN$ papers can be upgraded from $r_{\min}$ to $r_{\min}+1$; we choose these papers as a band of width $w=s$ centered at the expected acceptance percentile.

Let $\rho$ denote the overlap (precision) of the LLM-defined band: the fraction of papers in the LLM-selected band that fall in the true borderline region under the venue’s eventual outcomes.
Then the expected number of extra reviews that land on truly borderline papers is $\rho sN$.
Under a random baseline that selects $sN$ papers uniformly, the expected overlap with a true borderline set of size $sN$ is $s^2N$ papers.

Let $\delta_B$ denote the marginal flip rate for borderline papers and let $\delta_{\neg B}$ denote the marginal flip rate for non-borderline papers.
Define $\Delta := \delta_B - \delta_{\neg B}$ as the incremental decision-reliability gain from allocating an extra review to a borderline paper rather than a non-borderline paper (measured via a flip-sensitivity proxy in \S\ref{sec:estimation}, and ideally via randomized estimation in a pilot).
The expected number of net improved paper decisions is therefore $(\rho s - s^2)N\Delta.$ This expression is a first-order accounting: it translates overlap quality ($\rho$) and marginal review value ($\Delta$) into expected improvements under a fixed extra-review budget.
It abstracts away heterogeneity across papers, reviewer calibration differences, and topic-dependent variance; those effects would naturally lead to a distribution of gains rather than a single scalar.
We therefore interpret the formula as a transparent scale estimate rather than a precise causal prediction.

Using ICLR 2025 retrospective estimates (see \S\ref{sec:estimation}) with $\rho=\EstRho$ and $\Delta=\EstDelta$, and taking $N=\EstImpactN$ and $s=\EstImpactS$, the expected net improved paper decisions are about $\EstNetImpact$ corrected decisions.

\section{Estimating Parameters}
\label{sec:estimation}

We study a random sample of 1{,}000 ICLR 2025 submissions.
We run 40 rounds of random pairwise comparisons and fit a Bradley--Terry model.
This yields \EstRhoRounds~rounds $\times$ (\EstRhoPapers/2) matches per round, for a total of 20{,}000 pairwise battles.
For efficiency, we truncate each paper to at most the first 10 pages before LLM extraction.
In this 1,000-paper setting, the full LLM-based ranking cost about \$\EstRhoCostDollars{}.
We will release the paper list, extracted text, and LLM pairwise responses used in these experiments.

\subsection{Estimating the borderline overlap fraction $\rho$}
We estimate $\rho$ retrospectively on ICLR 2025 submissions with public outcomes.
We report the API model identifier logged during the runs (\EstRhoModel).
We run pairwise LLM comparisons (\EstRhoRounds~rounds over \EstRhoPapers~papers using \EstRhoModel) and fit a Bradley--Terry model to obtain an LLM ranking.
We define a proxy ``human ordering'' by decision tier (Reject $<$ Accept $<$ Spotlight $<$ Oral) and mean reviewer score within tier, acknowledging that this operationalizes the conference process rather than ground-truth quality.
This operationalizes the borderline set using observed outcomes and scores, so it reflects the review process rather than a latent ground-truth threshold.
We then define the borderline band as a quantile window centered at the borderline center (acceptance percentile).
Let $c$ denote the borderline center (acceptance percentile) and let $w$ denote the borderline band fraction.
The band spans the quantile interval $[c-w/2,\,c+w/2]$.
In our experiments we use $c=1-\TargetAcceptanceRateFrac$ and $w=\BorderlineFrac$.
We compute $\rho$ as the borderline overlap fraction of the LLM borderline set with respect to the human borderline set (overlap divided by the size of the LLM set).
Because we enforce $\lvert B_{\text{LLM}}\rvert=\lvert B_{\text{human}}\rvert$, this overlap is equal to both precision and recall; we report it as $\rho$ for simplicity.
\[
\rho = \frac{\lvert B_{\text{LLM}} \cap B_{\text{human}} \rvert}{\lvert B_{\text{LLM}} \rvert}.
\]
In practice, ACs and SACs may adjust assignments after the initial match; those adjustments can be applied on top of a targeted allocation policy.
We therefore compare the LLM-defined band to a random baseline (selecting $sN$ papers uniformly), which provides a neutral reference point for marginal review placement in the absence of any targeted policy.

\subsection{Estimating the marginal review effect $\Delta$}
We estimate $\Delta$ on the same \EstRhoPapers-paper ICLR 2025 sample, restricted to papers with $\ge 4$ reviews.
We use the same human ranking and borderline band as above.
For a paper with scores $s_1,\ldots,s_k$, we compute the mean $\mu$ and variance $\sigma^2$ of the $k$ scores.
We fit a logistic model $p=\operatorname{sigmoid}(\beta_0+\beta_1\mu+\beta_2\sigma^2)$ on observed decisions, where $p=P(\text{accept}\mid \mu,\sigma^2)$.
This calibrated flip model is an operational proxy and does not identify a causal effect of adding a review; a randomized extra-review design could estimate $\Delta$ causally.
For each review $i$, we form the leave-one-out statistics $(\mu_{-i},\sigma^2_{-i})$ and compute $p_{-i}$.
A \emph{flip} occurs if the accept indicator changes after removing a review: $\mathbf{1}\{p \ge 0.5\} \neq \mathbf{1}\{p_{-i} \ge 0.5\}$.
We aggregate flips over all leave-one-out trials within the borderline set and within the non-borderline set, and define $\Delta$ as the difference in flip rates.
Let $\delta_B$ denote the flip rate within the borderline set and let $\delta_{\neg B}$ denote the flip rate outside the borderline set.
\begin{align*}
\Delta &= \delta_B - \delta_{\neg B}, \\
\delta_B &= \frac{\#\text{ flips in }B}{\#\text{ trials in }B}, \\
\delta_{\neg B} &= \frac{\#\text{ flips in }\neg B}{\#\text{ trials in }\neg B}.
\end{align*}
We report a Wald 95\% confidence interval for $\Delta$ using a difference-in-proportions standard error,
\begin{align*}
\mathrm{SE}(\Delta) &= \sqrt{\frac{\delta_B(1-\delta_B)}{n_B}+\frac{\delta_{\neg B}(1-\delta_{\neg B})}{n_{\neg B}}}, \\
\mathrm{CI}_{0.95}(\Delta) &= \Delta\pm1.96\,\mathrm{SE}(\Delta),
\end{align*}
where $n_B$ and $n_{\neg B}$ are the numbers of leave-one-out trials in the borderline and non-borderline sets.
This yields an average $\Delta$ and likely masks heterogeneity across papers with different score variance and reviewer calibration.

\paragraph{Robustness and falsifiability.}
These estimates are retrospective proxies, so we report ablations and confidence intervals to assess stability.
If $\rho$ were close to the random baseline or if $\Delta$ were indistinguishable from zero on the 1{,}000-paper sample, the policy would have little expected benefit and would not justify deployment.
Conversely, persistent separation between the LLM ranking and the random baseline, together with a positive $\Delta$, supports the position that marginal reviews should be targeted rather than spread uniformly.

\section{Empirical Validation}
\label{sec:results}
The practical usefulness of our position hinges on empirical support: how many paper decisions are improved in expectation, and how accurately LLMs identify borderline papers.
We therefore treat this section as empirical validation aimed at understanding how impactful the position is in terms of expected net improved paper decisions, rather than as standalone algorithmic results.
Table~\ref{tab:est-results} summarizes the retrospective estimates from ICLR 2025 (\EstRhoPapers-paper sample, \EstRhoRounds~rounds with \EstRhoModel, \BorderlinePct~borderline band fraction).
The expected net improved paper decisions are computed as $(\rho s - s^2)N\Delta$ with $N=\EstImpactN$ and $s=\EstImpactS$.

\begin{table}[t]
  \centering
  \begin{tabular}{c c c}
    \hline
    $\rho$  & $\Delta$  & Expected net improved paper decisions \\
    \hline
    \EstRho & \EstDelta & \EstNetImpact                         \\
    \hline
  \end{tabular}
  \caption{ICLR 2025 retrospective estimates from \EstRhoPapers~papers (\EstRhoRounds~rounds, \EstRhoModel); borderline band fraction \BorderlinePct~around the acceptance percentile center; $\Delta$ computed on papers with $\ge 4$ reviews using the calibrated flip model.
    Expected net improved paper decisions are computed as $(\rho s - s^2)N\Delta$ with $N=\EstImpactN$ and $s=\EstImpactS$.
  }
  \label{tab:est-results}
\end{table}

\subsection{Ablations}
\begin{figure*}[t]
  \centering
  \begin{subfigure}[t]{0.48\textwidth}
    \centering
    \includegraphics[width=\linewidth]{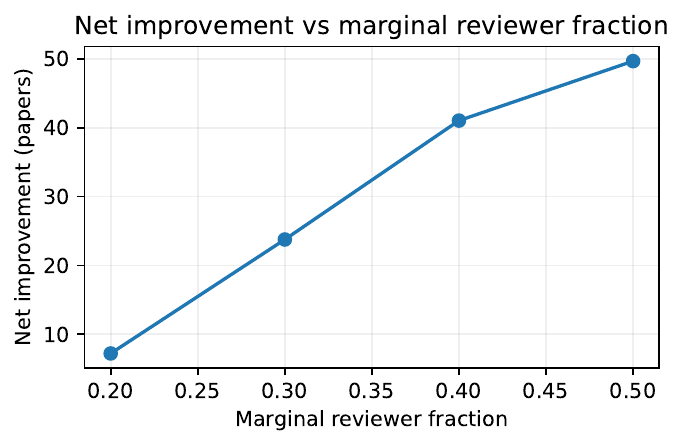}
    \caption{Expected net improved paper decisions sensitivity to the marginal reviewer fraction.
      We vary the marginal reviewer fraction and set the borderline band fraction to match it, then recompute $\rho$ and the expected net improved paper decisions under the review reallocation.
      Expected net improved paper decisions are shown relative to random allocation, using $(\rho s - s^2)N\Delta$ with $s$ fixed.
    }
    \label{fig:ablation-marginal}
  \end{subfigure}\hfill
  \begin{subfigure}[t]{0.48\textwidth}
    \centering
    \includegraphics[width=\linewidth]{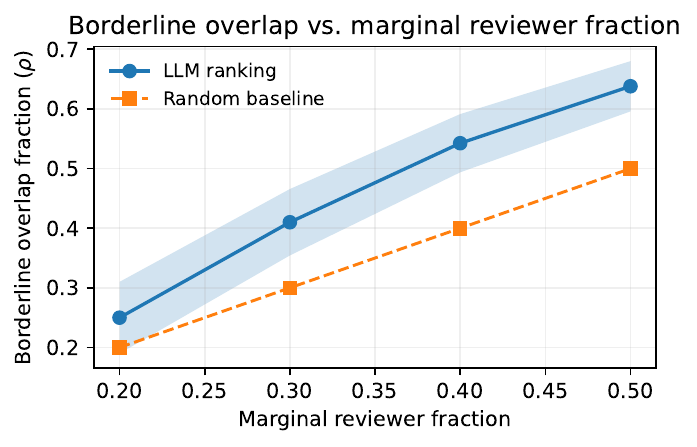}
    \caption{Borderline overlap fraction as a function of the marginal reviewer fraction.
      The dashed series shows the random-baseline $\rho$ implied by each band fraction, and shaded bands show the 95\% Wald confidence interval for $\rho$.
    }
    \label{fig:ablation-marginal-rho}
  \end{subfigure}
  \caption{Sensitivity to marginal reviewer fraction.}
  \label{fig:ablation-marginal-pair}
\end{figure*}

\begin{figure*}[t]
  \centering
  \begin{subfigure}[t]{0.48\textwidth}
    \centering
    \includegraphics[width=\linewidth]{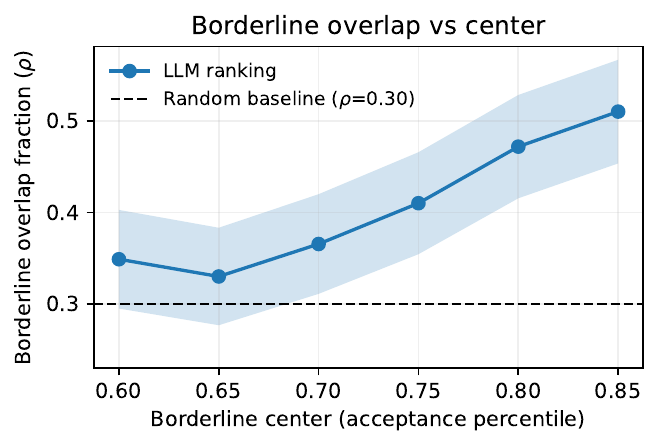}
    \caption{Centering ablation for the borderline center (acceptance percentile).
      We shift the band center and recompute $\rho$; the horizontal line shows the random-baseline $\rho$ implied by the \BorderlinePct~band fraction.
      Shaded bands show the 95\% Wald confidence interval for $\rho$ under the binomial overlap model.
    }
    \label{fig:ablation-centering}
  \end{subfigure}\hfill
  \begin{subfigure}[t]{0.48\textwidth}
    \centering
    \includegraphics[width=\linewidth]{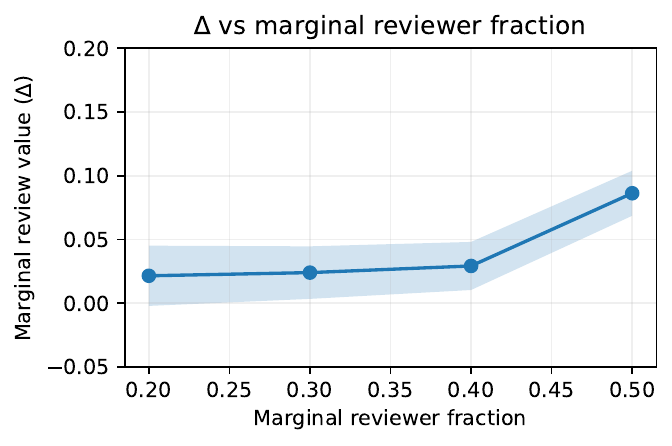}
    \caption{Marginal review value sensitivity to the marginal reviewer fraction.
      We vary the marginal reviewer fraction (and set the borderline band fraction to match it) and recompute $\Delta$ from leave-one-review counterfactuals under the calibrated flip rule.
    }
    \label{fig:ablation-delta-width}
  \end{subfigure}
  \caption{Sensitivity to centering and marginal review value.}
  \label{fig:ablation-centering-pair}
\end{figure*}

Using the Wald interval from Section~\ref{sec:estimation}, we obtain $\Delta=\EstDelta$ with 95\% CI $[\EstDeltaCiLow,\EstDeltaCiHigh]$.
We ablate the borderline band fraction and centering using cached LLM rankings and the calibrated $\Delta$ estimator.
For the centering ablation, we report 95\% confidence intervals for $\rho$ under a binomial overlap model with $m=\lvert B_{h}\rvert$ and $K=\lvert B_{h}\cap B_{\mathrm{LLM}}\rvert$.
We use the Wald interval
\begin{equation}
  \hat{\rho}=\frac{K}{m},\qquad
  \mathrm{CI}_{0.95}(\rho)=\hat{\rho}\pm1.96\sqrt{\frac{\hat{\rho}(1-\hat{\rho})}{m}}.
  \label{eq:rho-ci}
\end{equation}
For the band-fraction ablation, we set the borderline band fraction to match the marginal reviewer fraction $s$ and recompute $\rho$ or $\Delta$ under the new band definition.
We report all ablations on the \AblationPapers-paper sample with \AblationRounds~rounds.
We summarize the band-fraction and centering sensitivity of expected net improved paper decisions and $\rho$ in Figures~\ref{fig:ablation-marginal}--\ref{fig:ablation-centering}, including the random-baseline reference.
Figure~\ref{fig:ablation-delta-width} reports the corresponding sensitivity of $\Delta$ under the calibrated flip rule.
In Figure~\ref{fig:ablation-marginal}, expected net improved paper decisions increase with the marginal reviewer fraction; at $s=0.5$ this yields roughly 50 improved decisions in our setting.
Figure~\ref{fig:ablation-marginal-rho} shows that across a wide range of marginal reviewer fractions, the borderline overlap fraction remains statistically above the random baseline, indicating that the LLM ranking is informative for identifying borderline papers.
Figure~\ref{fig:ablation-centering} shows higher overlap as the borderline center moves to higher acceptance percentiles, consistent with the LLM being better at identifying top papers than separating near-cutoff papers.
Figure~\ref{fig:ablation-delta-width} indicates that the marginal value of an extra review is positive across a wide range of marginal reviewer fractions.

\paragraph{Summary.}
Across ablations, the LLM ranking consistently exceeds the random baseline in overlap, yielding positive expected net improved paper decisions under the allocation rule.
This suggests that the policy is not overly sensitive to a single hyperparameter choice, which is important for operational deployment.

\subsection{Ranking AUC}
\paragraph{Mann--Whitney AUC.}
We also report a global ranking metric based on the Mann--Whitney AUC between LLM scores and binary accept labels.
Let $\theta_i$ be the LLM score for paper $i$, and let $y_i \in \{0,1\}$ indicate accept vs.
\ reject.
Let $n_{+}$ be the number of accepted papers and $n_{-}$ be the number of rejected papers in the sample.
The AUC is
\begin{equation*}
  \mathrm{AUC}=\Pr(s_{i^{+}} > s_{i^{-}}) + \tfrac{1}{2}\Pr(s_{i^{+}} = s_{i^{-}}),
\end{equation*}
which is the probability that a randomly chosen accepted paper ranks above a randomly chosen rejected paper, with ties split evenly.
We estimate it from the finite sample by
\begin{equation*}
  \widehat{\mathrm{AUC}}=\frac{1}{n_{+}n_{-}}\sum_{i:y_i=1}\sum_{j:y_j=0}\Big[\mathbb{I}(\theta_i>\theta_j)+\tfrac{1}{2}\mathbb{I}(\theta_i=\theta_j)\Big].
\end{equation*}
Table~\ref{tab:auc} reports the Mann--Whitney AUC for the 1{,}000-paper run (random baseline AUC $=0.5$) and shows diminishing returns as the number of rounds increases.

\begin{table}[t]
  \centering
  \IfFileExists{\AucTablePath}{\begin{tabular}{lrrrr}
\toprule
Rounds & Papers & $n_{+}$ & $n_{-}$ & AUC \\
\midrule
40 & 1000 & 366 & 634 & 0.708 \\
30 & 1000 & 366 & 634 & 0.705 \\
20 & 1000 & 366 & 634 & 0.701 \\
10 & 1000 & 366 & 634 & 0.686 \\
5 & 1000 & 366 & 634 & 0.652 \\
\bottomrule
\end{tabular}
}{\textit{AUC table pending.}}
  \caption{Mann--Whitney AUC for the LLM ranking against binary accept labels.
    The table reports $n_{+}$ accepted papers and $n_{-}$ rejected papers in each sample.
  }
  \label{tab:auc}
\end{table}

We present ablations on model capability as well as full-paper vs abstract comparisons in the Appendix.

\section{Discussion}
\label{sec:discussion}

\begin{figure}
    \centering
    \includegraphics[width=0.45\linewidth]{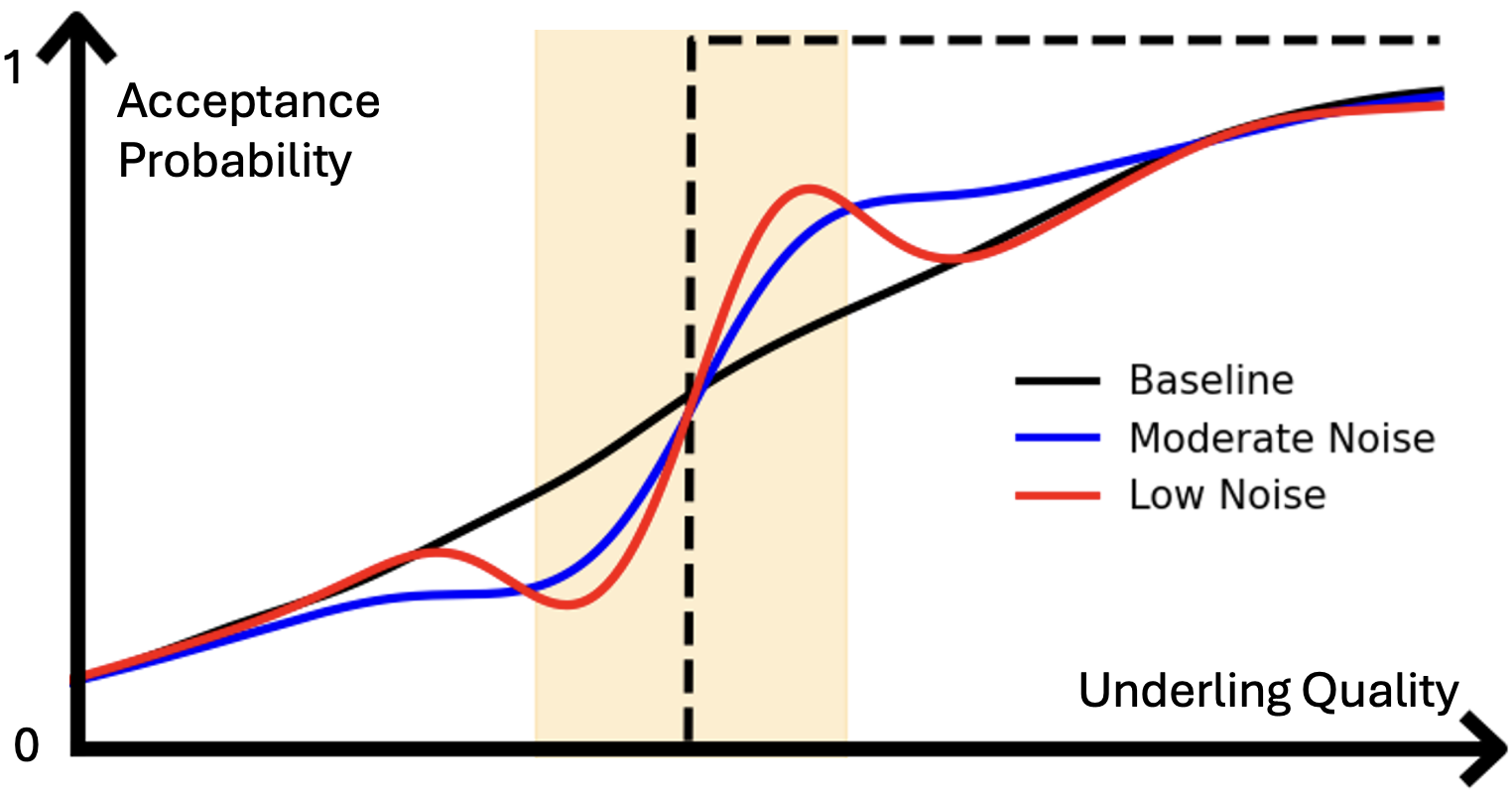}
    \caption{Toy Example of Acceptance Probability as a function of Underlying Quality: The black dotted line represents the switch from 0 probability to 1 probability of acceptance when the quality exceeds a certain threshold (target quantile). This is the first best behavior that could only be achieved when the underlying quality is known to us. Reviewer feedback is a noisy signal of the quality, leading to the black curve. Although having quality exceeding the threshold may no longer guarantee acceptance, higher quality leads to higher acceptance probability. The red curve represents the acceptance probability under our scheme when LLM has extremely high fidelity: if an author has an option to reduce one's quality, they could intentionally reduce their quality to increase the acceptance probability. However, we can see that, when an LLM has moderate noise, the acceptance probability follows the blue curve, which is monotone, rendering gaming impossible.}
    \label{fig:curve}
\end{figure}

\textbf{Gaming.}
Authors might try to steer their paper toward the borderline band to attract an extra review: If the paper is originally good, extra review reduces noise, thereby increasing acceptance chance. An originally not as good paper can also reduce its quality to steer away from getting an extra reviewer. However, we suspect that this gaming is impossible in practice since it requires knowledge about others' qualities.\footnote{This is usually feasible only with continuous observation~\cite{edelman2007internet} or continuum of agents with a known distribution~\cite{sornwanee20251}. Both conditions are not met in a practical peer review setting.} LLM ranking is comparative and noisy, further reducing the incentive of downgrading paper (See figure~\ref{fig:curve}). We also recommend allocating only a fraction of extra reviews via the LLM signal, with the remainder assigned uniformly at random as a decoy, so that receiving a fourth review is not a clear signal of borderline status.

\

\section{Alternative Views}
\label{sec:alternative-views}
\textbf{Decision accuracy is not the right objective.}
Some may argue that the objective should be to maximize the chance that the very best papers are accepted.
That would favor allocating extra reviews to the top-ranked papers rather than to the borderline band.
Others may argue the opposite, that weaker papers deserve more reviews to receive helpful feedback.
We see these as different objectives at different stages.
Identifying the very best papers can be prioritized later, for example when area chairs and senior area chairs deliberate on awards, talks, or top-presentation slots, which typically occur after accept or reject outcomes are fixed and need not respect tight resubmission timelines.
By contrast, accept or reject decisions must be timely, and they are most sensitive near the boundary, so reallocating marginal reviews there yields the highest expected improvement in correct decisions.
On the low end, additional reviews may be less informative for outcomes and can be redundant when multiple reviewers already agree; structured feedback and mentoring can be provided without changing the review-count allocation.
Our position is narrower: given a fixed extra-review budget for acceptance decisions, we target the region where marginal reviews are most likely to change accept or reject outcomes.

\textbf{Avoid LLMs in reviewing.}
Concerns about LLM review reliability are grounded in recent evaluations and surveys \citep{akella2025prereviewpeerreviewpitfalls,zhou-etal-2024-llm,10.1016/j.inffus.2025.103332}.
Another view is that any LLM involvement introduces unacceptable bias.
Recent work reports systematic differences in LLM rankings across paper categories, which could skew where extra reviews are placed.
This concern is strongest when LLMs are used to make accept or reject decisions, because any bias can directly change outcomes and creates incentives for authors to optimize for the prompt rather than scientific impact.
Our design uses LLMs only to allocate review \emph{counts}, keeps the signal hidden from decision-makers, and preserves fully human outcomes.
As a result, any bias mainly affects where additional scrutiny is applied, not the decision rule itself, and the incentive to game the LLM is sharply reduced.
We also recommend allocating only a fraction of extra reviews via the LLM signal, with the remainder assigned uniformly at random as a decoy, so that four reviews do not reliably indicate a paper is borderline.
These mitigations do not eliminate bias risk, but they reduce the chance that LLM outputs systematically tilt final decisions.

\textbf{Time and cost.}
One could argue that improving a modest number of decisions is not worth the cost of running an LLM ranking.
In our 1,000-paper run, the full ranking cost about \$\EstRhoCostDollars{}.
Scaling linearly to a NeurIPS-scale full run ($N=\EstImpactN$) yields about \$\CostScaledDollars{} before batch discounts (about \$\CostScaledDollarsAfterDiscount{} with a \CostBatchDiscount{} discount), holding prompt lengths and pricing fixed.
Dividing by our expected net improved paper decisions (Section~\ref{sec:results}) yields an implied cost per improved decision of about \$75 (using \CostScaledDollarsAfterDiscount{} and \EstNetImpact{}).
Given the total time invested by expert reviewers and the reputational cost of errors at scale, this cost may be reasonable.
Operationally, the ranking can run in parallel with reviewer bidding, so it need not add elapsed time to the review cycle.

\section{Conclusion}
\label{sec:conclusion}
We argue that reallocating spare reviewer capacity toward a statistically defined borderline band can improve decision quality without changing reviewer load.
Our empirical estimates of the borderline overlap fraction $\rho$ and the marginal review value $\Delta$, together with ablations, indicate that the gains are robust to reasonable variations in band fraction and centering.
More broadly, this position advocates using LLMs to guide \emph{allocation} rather than \emph{judgment}: the goal is to focus scarce human attention where it is most likely to change outcomes while keeping final decisions entirely human.
Operationally, this is a minimal change to current reviewing workflows; with explicit guardrails, it carries low risk of bias amplification and low incentives for gaming.

\bibliography{main}

\begin{thebibliography}{36}
\providecommand{\natexlab}[1]{#1}
\providecommand{\url}[1]{\texttt{#1}}
\expandafter\ifx\csname urlstyle\endcsname\relax
  \providecommand{\doi}[1]{doi: #1}\else
  \providecommand{\doi}{doi: \begingroup \urlstyle{rm}\Url}\fi

\bibitem[{AAAI}(2025)]{aaai2026llmpressrelease}
{AAAI}.
\newblock Association for the advancement of artificial intelligence launches
  ai-powered peer review assessment system.
\newblock Press release, 2025.
\newblock URL
  \url{https://aaai.org/wp-content/uploads/2025/05/AAAI-LLM-Press-Release.pdf}.
\newblock Accessed 2026-01-28.

\bibitem[Akella et~al.(2025)Akella, Siravuri, and
  Rohatgi]{akella2025prereviewpeerreviewpitfalls}
Akhil~Pandey Akella, Harish~Varma Siravuri, and Shaurya Rohatgi.
\newblock Pre-review to peer review: Pitfalls of automating reviews using large
  language models, 2025.
\newblock URL \url{https://arxiv.org/abs/2512.22145}.

\bibitem[Bradley and Terry(1952)]{bradley1952rank}
Ralph~Allan Bradley and Milton~E. Terry.
\newblock Rank analysis of incomplete block designs: I. the method of paired
  comparisons.
\newblock \emph{Biometrika}, 39\penalty0 (3/4):\penalty0 324--345, 1952.

\bibitem[Charlin and Zemel(2013)]{Charlin2013TheTP}
Laurent Charlin and Richard~S. Zemel.
\newblock The toronto paper matching system: An automated paper-reviewer
  assignment system, 2013.
\newblock URL \url{https://api.semanticscholar.org/CorpusID:680003}.

\bibitem[Charlin et~al.(2012)Charlin, Zemel, and
  Boutilier]{charlin2012frameworkoptimizingpapermatching}
Laurent Charlin, Richard~S. Zemel, and Craig Boutilier.
\newblock A framework for optimizing paper matching, 2012.
\newblock URL \url{https://arxiv.org/abs/1202.3706}.

\bibitem[Ebrahimi et~al.(2025)Ebrahimi, Sadeghian, Ghorbanpour, Arabzadeh,
  Salamat, Li, Le, Bashari, and Bagheri]{10.1145/3746252.3761506}
Sajad Ebrahimi, Soroush Sadeghian, Ali Ghorbanpour, Negar Arabzadeh, Sara
  Salamat, Muhan Li, Hai~Son Le, Mahdi Bashari, and Ebrahim Bagheri.
\newblock Rottenreviews: Benchmarking review quality with human and llm-based
  judgments.
\newblock In \emph{Proceedings of the 34th ACM International Conference on
  Information and Knowledge Management}, CIKM '25, page 5642–5649, New York,
  NY, USA, 2025. Association for Computing Machinery.
\newblock ISBN 9798400720406.
\newblock \doi{10.1145/3746252.3761506}.
\newblock URL \url{https://doi.org/10.1145/3746252.3761506}.

\bibitem[Edelman et~al.(2007)Edelman, Ostrovsky, and
  Schwarz]{edelman2007internet}
Benjamin Edelman, Michael Ostrovsky, and Michael Schwarz.
\newblock Internet advertising and the generalized second-price auction:
  Selling billions of dollars worth of keywords.
\newblock \emph{American economic review}, 97\penalty0 (1):\penalty0 242--259,
  2007.

\bibitem[Epstein et~al.(2026)Epstein, Winnicki, Sornwanee, and
  Dwaraknath]{epstein2026llms}
Elliot~L Epstein, John Winnicki, Thanawat Sornwanee, and Rajat~Vadiraj
  Dwaraknath.
\newblock {LLM}s are overconfident: Evaluating confidence interval calibration
  with fermieval.
\newblock In \emph{AAAI 2026 Workshop on Assessing and Improving Reliability of
  Foundation Models in the Real World}, 2026.
\newblock URL \url{https://openreview.net/forum?id=yUyFITL0wv}.

\bibitem[Francois(2015)]{francois2015arbitrarinesspeerreviewbayesian}
Olivier Francois.
\newblock Arbitrariness of peer review: A bayesian analysis of the nips
  experiment, 2015.
\newblock URL \url{https://arxiv.org/abs/1507.06411}.

\bibitem[{Gemini Team}(2025)]{geminiteam2025geminifamilyhighlycapable}
{Gemini Team}.
\newblock Gemini: A family of highly capable multimodal models, 2025.
\newblock URL \url{https://arxiv.org/abs/2312.11805}.

\bibitem[Idahl and Ahmadi(2025)]{idahl2025openreviewerspecializedlargelanguage}
Maximilian Idahl and Zahra Ahmadi.
\newblock Openreviewer: A specialized large language model for generating
  critical scientific paper reviews, 2025.
\newblock URL \url{https://arxiv.org/abs/2412.11948}.

\bibitem[Jecmen et~al.(2020)Jecmen, Zhang, Liu, Shah, Conitzer, and
  Fang]{NEURIPS2020_93fb3947}
Steven Jecmen, Hanrui Zhang, Ryan Liu, Nihar Shah, Vincent Conitzer, and Fei
  Fang.
\newblock Mitigating manipulation in peer review via randomized reviewer
  assignments.
\newblock In H.~Larochelle, M.~Ranzato, R.~Hadsell, M.F. Balcan, and H.~Lin,
  editors, \emph{Advances in Neural Information Processing Systems}, volume~33,
  pages 12533--12545. Curran Associates, Inc., 2020.
\newblock URL
  \url{https://proceedings.neurips.cc/paper_files/paper/2020/file/93fb39474c51b8a82a68413e2a5ae17a-Paper.pdf}.

\bibitem[Kobren et~al.(2019)Kobren, Saha, and
  McCallum]{kobren2019papermatchinglocalfairness}
Ari Kobren, Barna Saha, and Andrew McCallum.
\newblock Paper matching with local fairness constraints, 2019.
\newblock URL \url{https://arxiv.org/abs/1905.11924}.

\bibitem[Leyton-Brown et~al.(2022)Leyton-Brown, Mausam, Nandwani, Zarkoob,
  Cameron, Newman, and Raghu]{leytonbrown2022matchingpapersreviewerslarge}
Kevin Leyton-Brown, Mausam, Yatin Nandwani, Hedayat Zarkoob, Chris Cameron,
  Neil Newman, and Dinesh Raghu.
\newblock Matching papers and reviewers at large conferences, 2022.
\newblock URL \url{https://arxiv.org/abs/2202.12273}.

\bibitem[Leyton-Brown et~al.(2024)Leyton-Brown, Mausam, Nandwani, Zarkoob,
  Cameron, Newman, and Raghu]{LEYTONBROWN2024104119}
Kevin Leyton-Brown, Mausam, Yatin Nandwani, Hedayat Zarkoob, Chris Cameron,
  Neil Newman, and Dinesh Raghu.
\newblock Matching papers and reviewers at large conferences.
\newblock \emph{Artificial Intelligence}, 331:\penalty0 104119, 2024.
\newblock ISSN 0004-3702.
\newblock \doi{https://doi.org/10.1016/j.artint.2024.104119}.
\newblock URL
  \url{https://www.sciencedirect.com/science/article/pii/S0004370224000559}.

\bibitem[Liang et~al.(2023)Liang, Zhang, Cao, Wang, Ding, Yang, Vodrahalli, He,
  Smith, Yin, McFarland, and Zou]{liang2023largelanguagemodelsprovide}
Weixin Liang, Yuhui Zhang, Hancheng Cao, Binglu Wang, Daisy Ding, Xinyu Yang,
  Kailas Vodrahalli, Siyu He, Daniel Smith, Yian Yin, Daniel McFarland, and
  James Zou.
\newblock Can large language models provide useful feedback on research papers?
  a large-scale empirical analysis, 2023.
\newblock URL \url{https://arxiv.org/abs/2310.01783}.

\bibitem[Liu and Shah(2023)]{liu2023reviewergptexploratorystudyusing}
Ryan Liu and Nihar~B. Shah.
\newblock Reviewergpt? an exploratory study on using large language models for
  paper reviewing, 2023.
\newblock URL \url{https://arxiv.org/abs/2306.00622}.

\bibitem[Liu et~al.(2014)Liu, Suel, and Memon]{10.1145/2645710.2645749}
Xiang Liu, Torsten Suel, and Nasir Memon.
\newblock A robust model for paper reviewer assignment.
\newblock In \emph{Proceedings of the 8th ACM Conference on Recommender
  Systems}, RecSys '14, page 25–32, New York, NY, USA, 2014. Association for
  Computing Machinery.
\newblock ISBN 9781450326681.
\newblock \doi{10.1145/2645710.2645749}.
\newblock URL \url{https://doi.org/10.1145/2645710.2645749}.

\bibitem[Lu and Kong(2023)]{lu2023calibrating}
Yuxuan Lu and Yuqing Kong.
\newblock Calibrating {\textquotedblleft}cheap signals{\textquotedblright} in
  peer review without a prior.
\newblock In \emph{Thirty-seventh Conference on Neural Information Processing
  Systems}, 2023.
\newblock URL \url{https://openreview.net/forum?id=xr3KAzboHY}.

\bibitem[Plank and van Dalen(2019)]{Plank2019CiteTrackedAL}
Barbara Plank and Reinard van Dalen.
\newblock Citetracked: A longitudinal dataset of peer reviews and citations.
\newblock In \emph{BIRNDL@SIGIR}, 2019.
\newblock URL \url{https://api.semanticscholar.org/CorpusID:198489688}.

\bibitem[Robertson(2023)]{robertson2023gpt4slightlyhelpfulpeerreview}
Zachary Robertson.
\newblock Gpt4 is slightly helpful for peer-review assistance: A pilot study,
  2023.
\newblock URL \url{https://arxiv.org/abs/2307.05492}.

\bibitem[Saveski et~al.(2023)Saveski, Jecmen, Shah, and
  Ugander]{saveski2023counterfactual}
Martin Saveski, Steven Jecmen, Nihar~B Shah, and Johan Ugander.
\newblock Counterfactual evaluation of peer-review assignment policies.
\newblock In \emph{Thirty-seventh Conference on Neural Information Processing
  Systems}, 2023.
\newblock URL \url{https://openreview.net/forum?id=rhIfzCZoXG}.

\bibitem[Shah et~al.(2018)Shah, Tabibian, Muandet, Guyon, and von
  Luxburg]{shah2018designanalysisnips2016}
Nihar~B. Shah, Behzad Tabibian, Krikamol Muandet, Isabelle Guyon, and Ulrike
  von Luxburg.
\newblock Design and analysis of the nips 2016 review process, 2018.
\newblock URL \url{https://arxiv.org/abs/1708.09794}.

\bibitem[Sornwanee(2025)]{sornwanee20251}
Thanawat Sornwanee.
\newblock 1-dimensional normal competitive market equilibrium.
\newblock 2025.
\newblock URL \url{https://arxiv.org/abs/2505.08425}.

\bibitem[Srinivasan and
  Morgenstern(2023)]{srinivasan2023auctionspeerpredictionacademic}
Siddarth Srinivasan and Jamie Morgenstern.
\newblock Auctions and peer prediction for academic peer review, 2023.
\newblock URL \url{https://arxiv.org/abs/2109.00923}.

\bibitem[Stelmakh et~al.(2019)Stelmakh, Shah, and
  Singh]{stelmakh2019peerreview4allfairaccuratereviewer}
Ivan Stelmakh, Nihar~B. Shah, and Aarti Singh.
\newblock Peerreview4all: Fair and accurate reviewer assignment in peer review,
  2019.
\newblock URL \url{https://arxiv.org/abs/1806.06237}.

\bibitem[Su et~al.(2025)Su, Zhang, Collina, Yan, Li, Cho, Fan, Roth, and
  Su]{Su17072025}
Buxin Su, Jiayao Zhang, Natalie Collina, Yuling Yan, Didong Li, Kyunghyun Cho,
  Jianqing Fan, Aaron Roth, and Weijie Su.
\newblock The icml 2023 ranking experiment: Examining author self-assessment in
  ml/ai peer review.
\newblock \emph{Journal of the American Statistical Association}, 0\penalty0
  (0):\penalty0 1--12, 2025.
\newblock \doi{10.1080/01621459.2025.2510006}.
\newblock URL \url{https://doi.org/10.1080/01621459.2025.2510006}.

\bibitem[Su(2021)]{su2021you}
Weijie~J Su.
\newblock You are the best reviewer of your own papers: An owner-assisted
  scoring mechanism.
\newblock In A.~Beygelzimer, Y.~Dauphin, P.~Liang, and J.~Wortman Vaughan,
  editors, \emph{Advances in Neural Information Processing Systems}, 2021.
\newblock URL \url{https://openreview.net/forum?id=xmx5rE9QP7R}.

\bibitem[Xu et~al.(2023)Xu, Jecmen, Song, and Fang]{xu2023a}
Yixuan~Even Xu, Steven Jecmen, Zimeng Song, and Fei Fang.
\newblock A one-size-fits-all approach to improving randomness in paper
  assignment.
\newblock In \emph{Thirty-seventh Conference on Neural Information Processing
  Systems}, 2023.
\newblock URL \url{https://openreview.net/forum?id=D94QKZA7UP}.

\bibitem[Ye et~al.(2024)Ye, Pang, Chai, Chen, Yin, Xiang, Dong, Shao, and
  Chen]{ye2024yetrevealingrisksutilizing}
Rui Ye, Xianghe Pang, Jingyi Chai, Jiaao Chen, Zhenfei Yin, Zhen Xiang, Xiaowen
  Dong, Jing Shao, and Siheng Chen.
\newblock Are we there yet? revealing the risks of utilizing large language
  models in scholarly peer review, 2024.
\newblock URL \url{https://arxiv.org/abs/2412.01708}.

\bibitem[Zhang and Abernethy(2025)]{zhang2025reviewingscientificpaperscritical}
Tianmai~M. Zhang and Neil~F. Abernethy.
\newblock Reviewing scientific papers for critical problems with reasoning
  llms: Baseline approaches and automatic evaluation, 2025.
\newblock URL \url{https://arxiv.org/abs/2505.23824}.

\bibitem[Zhang et~al.(2025)Zhang, ZHANG, Ji, Hua, Haber, Cao, and
  Liang]{zhang2025from}
Yaohui Zhang, Haijing ZHANG, Wenlong Ji, Tianyu Hua, Nick Haber, Hancheng Cao,
  and Weixin Liang.
\newblock From replication to redesign: Exploring pairwise comparisons for
  {LLM}-based peer review.
\newblock In \emph{The Thirty-ninth Annual Conference on Neural Information
  Processing Systems}, 2025.
\newblock URL \url{https://openreview.net/forum?id=z5KTxW5sJd}.

\bibitem[Zhao et~al.(2025)Zhao, Tian, Xing, Zhang, Li, Qian, Cheng, and
  Li]{zhao2025naipv2debiasedpairwiselearning}
Penghai Zhao, Jinyu Tian, Qinghua Xing, Xin Zhang, Zheng Li, Jianjun Qian,
  Ming-Ming Cheng, and Xiang Li.
\newblock Naipv2: Debiased pairwise learning for efficient paper quality
  estimation, 2025.
\newblock URL \url{https://arxiv.org/abs/2509.25179}.

\bibitem[Zhou et~al.(2024)Zhou, Chen, and Yu]{zhou-etal-2024-llm}
Ruiyang Zhou, Lu~Chen, and Kai Yu.
\newblock Is {LLM} a reliable reviewer? a comprehensive evaluation of {LLM} on
  automatic paper reviewing tasks.
\newblock In Nicoletta Calzolari, Min-Yen Kan, Veronique Hoste, Alessandro
  Lenci, Sakriani Sakti, and Nianwen Xue, editors, \emph{Proceedings of the
  2024 Joint International Conference on Computational Linguistics, Language
  Resources and Evaluation (LREC-COLING 2024)}, pages 9340--9351, Torino,
  Italia, May 2024. ELRA and ICCL.
\newblock URL \url{https://aclanthology.org/2024.lrec-main.816/}.

\bibitem[Zhu et~al.(2025)Zhu, Weng, Yang, and Zhang]{zhu-etal-2025-deepreview}
Minjun Zhu, Yixuan Weng, Linyi Yang, and Yue Zhang.
\newblock {D}eep{R}eview: Improving {LLM}-based paper review with human-like
  deep thinking process.
\newblock In Wanxiang Che, Joyce Nabende, Ekaterina Shutova, and Mohammad~Taher
  Pilehvar, editors, \emph{Proceedings of the 63rd Annual Meeting of the
  Association for Computational Linguistics (Volume 1: Long Papers)}, pages
  29330--29355, Vienna, Austria, July 2025. Association for Computational
  Linguistics.
\newblock ISBN 979-8-89176-251-0.
\newblock \doi{10.18653/v1/2025.acl-long.1420}.
\newblock URL \url{https://aclanthology.org/2025.acl-long.1420/}.

\bibitem[Zhuang et~al.(2025)Zhuang, Chen, Xu, Jiang, and
  Lin]{10.1016/j.inffus.2025.103332}
Zhenzhen Zhuang, Jiandong Chen, Hongfeng Xu, Yuwen Jiang, and Jialiang Lin.
\newblock Large language models for automated scholarly paper review: A survey.
\newblock \emph{Inf. Fusion}, 124\penalty0 (C), December 2025.
\newblock ISSN 1566-2535.
\newblock \doi{10.1016/j.inffus.2025.103332}.
\newblock URL \url{https://doi.org/10.1016/j.inffus.2025.103332}.

\end{thebibliography}
\bibliographystyle{plainnat}

\appendix
\section{Additional Empirical Validation}
\subsection{Full paper vs.\ abstract-only inputs.}
\begin{figure*}[t]
  \centering
  \begin{subfigure}[t]{0.48\textwidth}
    \centering
    \includegraphics[width=\linewidth]{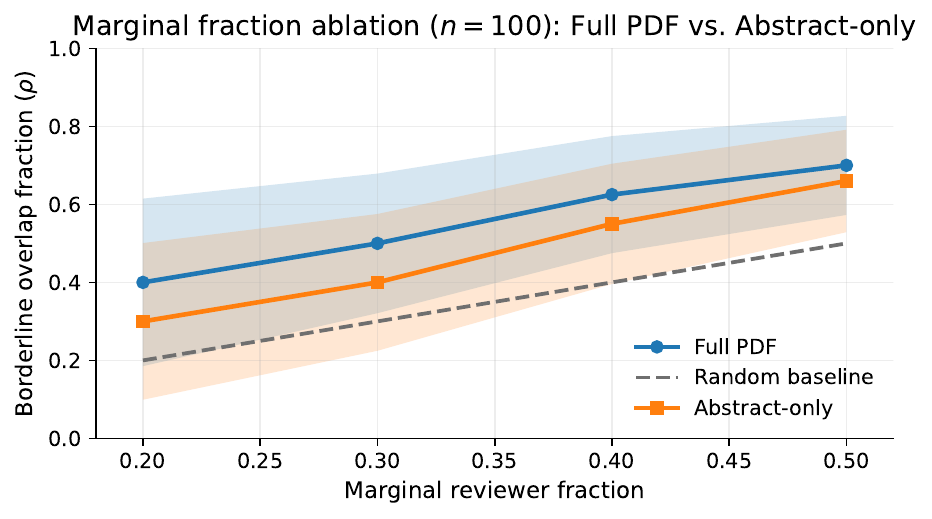}
    \caption{Marginal fraction ablation (\(n=100\)): full PDF vs.\ abstract-only.
      We report agreement \(\rho\) between the LLM judge and human judgments as a function of the marginal reviewer fraction when the judge compares either the full paper text or only the abstract; the dashed line shows the random baseline (shaded bands indicate uncertainty).
    }
    \label{fig:full-vs-abstract-rho}
  \end{subfigure}\hfill
  \begin{subfigure}[t]{0.48\textwidth}
    \centering
    \includegraphics[width=\linewidth]{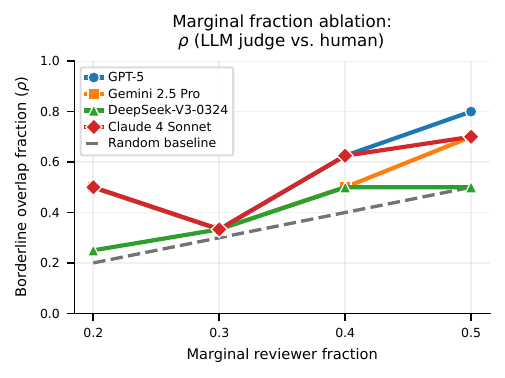}
    \caption{Model capability ablation.
      Borderline overlap fraction \(\rho\) (LLM judge vs.
      \ human) as a function of marginal reviewer fraction.
      In this small run, higher-capability judges tend to achieve higher agreement, while weaker judges track closer to the random baseline (dashed).
    }
    \label{fig:model-ablation-summary}
  \end{subfigure}
  \caption{Ablations on information access and judge capability.}
  \label{fig:appendix-ablations}
\end{figure*}

We ablate how much paper content the LLM judge sees by running the marginal-fraction ablation with either the full PDF text or only the abstract across 100 papers with 5 rounds.
Figure~\ref{fig:full-vs-abstract-rho} shows that both settings improve with larger marginal fractions and both outperform the random baseline, but access to the full paper yields consistently higher agreement with humans.
Concretely, \(\rho\) increases from \(\approx 0.40\) to \(\approx 0.70\) (full PDF) as the marginal reviewer fraction rises from 0.2 to 0.5, while abstract-only rises from \(\approx 0.30\) to \(\approx 0.66\), leaving a persistent (though narrowing) absolute gap of roughly \(0.04\text{--}0.10\).

This ablation highlights a cost--accuracy tradeoff.
Abstracts contain a large share of the signal needed for pairwise comparisons, but the full text provides additional evidence that measurably improves fidelity, especially at smaller marginal fractions.
From a practical standpoint, this matters because (as discussed in our Alternative Views cost analysis) total dollar cost and latency are dominated by input tokens, and full-PDF comparisons are materially more expensive than abstract-only prompts.
A natural compromise is a staged pipeline: run abstract-only comparisons broadly, and reserve full-PDF comparisons for papers near the estimated borderline where the extra signal is most valuable.

\subsection{Model capability ablation.}
We ablate the choice of LLM judge and measure agreement with human judgments using the borderline overlap fraction \(\rho\) (Figure~\ref{fig:model-ablation-summary}) in a small 20-paper, 30-round run.
Across the marginal fractions shown, the higher-capability judges tend to achieve higher agreement, with differences most apparent at larger marginal reviewer fractions.
At the largest fraction we evaluate (\(0.5\)), GPT-5 reaches \(\rho \approx 0.8\), Gemini 2.5 Pro and Claude 4 Sonnet are around \(\rho \approx 0.7\), while DeepSeek-V3-0324 is closer to the random baseline trend (around \(\rho \approx 0.5\)).
At smaller marginal fractions the curves are closer together and not strictly monotonic, which is plausible given the small scale of this test.

Overall, this ablation is intended as a lightweight sanity check (not a definitive ranking), but it supports the practical takeaway that the choice of judge can meaningfully affect how closely the evaluation tracks human decisions, especially when the marginal fraction is larger.

\end{document}